# Classifying superconductivity in ThH-ThD superhydrides/superdeuterides

E. F. Talantsev[1,2,*] and R. C. Mataira[3]

[1]M.N. Mikheev Institute of Metal Physics, Ural Branch, Russian Academy of Sciences, 18, S. Kovalevskoy St., Ekaterinburg, 620108, Russia

[2]NANOTECH Centre, Ural Federal University, 19 Mira St., Ekaterinburg, 620002, Russia

[3]Robinson Research Institute, Victoria University of Wellington, 69 Gracefield Rd., Lower Hutt, 5010, New Zealand

[*]E-mail: evgeny.talantsev@imp.uran.ru

*Abstract*

Satterthwaite and Toepke (1970 *Phys. Rev. Lett.* **25** 741) discovered that $Th_4H_{15}$-$Th_4D_{15}$ superhydrides are superconducting but exhibit no isotope effect. As the isotope effect is a fundamental prediction of electron-phonon mediated superconductivity described by Bardeen, Cooper, and Schrieffer (BCS) its absence alludes to some other mechanism. Soon after this work, Stritzker and Buckel (1972 *Zeitschrift für Physik A Hadrons and nuclei* **257** 1-8) reported that superconductors in the $PdH_x$-$PdD_x$ system exhibit the reverse isotope effect. Yussouff *et al* (1995 *Solid State Communications* **94** 549) extended this finding in $PdH_x$-$PdD_x$-$PdT_x$ systems. Renewed interest in hydrogen- and deuterium-rich superconductors is driven by the discovery of near-room-temperature superconductivity in highly-compressed $H_3S$ (Drozdov *et al.* 2015 *Nature* **525** 73) and $LaH_{10}$ (Somayazulu *et al* 2019 *Phys. Rev. Lett.* **122** 027001. Here we attempt to reaffirm or disprove our primary idea that the mechanism for near-room-temperature superconductivity in hydrogen-rich superconductors is not BCS electron-phonon mediated. To that end, we analyse the upper critical field data, $B_{c2}(T)$, in $Th_4H_{15}$-$Th_4D_{15}$ (Satterthwaite and Toepke 1970 *Phys. Rev. Lett.* **25** 741) as well as two recently discovered high-pressure hydrogen-rich phases of $ThH_9$ and $ThH_{10}$ (Semenok *et al* 2019 *Materials Today*, DOI: 10.1016/j.mattod.2019.10.005). We conclude that all known thorium super-hydrides/deuterides, to date, are unconventional superconductors – along with the heavy fermions, fullerenes, pnictides, cuprates – where we find they have $T_c/T_F$ ratios within a range of $0.008 < T_c/T_F < 0.120$, where $T_c$ is the superconducting transition temperature and $T_F$ is the Fermi temperature.

I. **Introduction**

The discovery of near-room-temperature (NRT) superconductivity in highly-compressed $H_3S$ ($T_c = 203$ K) [1], and the following discovery of superconductivity in $LaH_{10}$ ($T_c = 250$ K, $P = 150$ GPa) [2] (current status of the research in the field can be found elsewhere [3-7]), is widely held [8] as a success of the predictions of Ashcroft [9] and Ginzburg [10]. These predictions were based on electron-phonon interactions of Bardeen, Cooper, and Schrieffer's (BCS) theory of superconductivity [11]. This prediction, of NRT $T_c$ hydrides under pressure, and its subsequent discovery in $H_3S$ and $LaH_{10}$ were taken as affirmations that these systems were indeed conventional (BCS electron-phonon mediated) superconductors. However, as we have shown previously [12,13] the available data for these superhydrides can be successfully interpreted in the phenomenology of unconventional (non-BCS) superconductivity – suggesting that the mechanism is not BCS electron-phonon coupling. To further our analysis, and hopefully reiterate the need for new experimental data on the $H_3S$-$D_3S$ system, we revisit the thorium-based hydrides $Th_4H_{15}$-$Th_4D_{15}$, $ThH_9$, and $ThH_{10}$, to see if a similar conclusion has been overlooked.

The isotope effect in Bardeen-Cooper-Schrieffer (BCS) theory of superconductivity can be expressed in the form:

$$T_c \cdot M^\beta = const, \qquad (1)$$

where $M$ is isotope mass, and $\beta \approx 1/2$ (for weak-coupling limit of BCS theory [11]), is an indispensable feature of electron-phonon mediated superconductivity [1,11]. This effect was observed in several elemental superconductors, but not in all of them [14,15]. Geballe *et al* [16] were the first to find the absence of the isotope effect in ruthenium (more details can be found elsewhere [14,15]). Later, Satterthwaite and Toepke [17] reported the absence of the isotope effect in $Th_4H_{15}$-$Th_4D_{15}$ super-hydride/deuteride phases. Soon after [17], Stritzker and Buckel [18] experimentally found that the isotope effect in the palladium-hydrogen-deuterium ($PdH_x$-$PdD_x$) system has the opposite sign (the reverse isotope effect). Yussouff *et al* [19] extended this discovery to the full palladium-hydrogen-deuterium-tritium ($PdH_x$-$PdD_x$-$PdT_x$) system. This reverse isotope effect in the $PdH_x$-$PdD_x$-$PdT_x$ system is currently the subject of wide discussion [20,21]. As for the thorium hydrides/deuteride systems considered herein, detailed studies by Caton and Satterthwaite [22] reported a reverse isotope effect in $Th_4H_{15}$ $Th_4D_{15}$.

Discovery of NRT superconductivity in $H_3S$-$D_3S$ [1] and $LaH_{10}$ [2] has reinvigorated interest in the isotope effect in the superconducting hydrides/deuterides. It should be stressed that recent experimental results reported by Drozdov *et al.* [23] show that La-H and La-D NRT phases have different stoichiometry, i.e. $LaH_{10}$ vs $LaD_{11}$/$LaD_{12}$, and, thus, more experimental and theoretical studies are demanded to reveal the effect of the isotope effect on $T_c$ in LaH-LaD system, which should be separated from the effect of different chemical stoichiometry on $T_c$ in these superhydrides/superdeuterides.

These studies will support/disprove our previous proposal that hydrogen-rich compounds ($PdH_x$, $H_3S$, $LaH_{10}$) are unconventional superconductors [12,13] and, thus, the superconductivity in these compounds is not related to electron-phonon interaction. We should note that, so far, we have not included the following in our analysis or proposals:

1. Highly compressed silane $SiH_4$, first discovered by Eremets group, $T_c = 17\ K$ (observed at pressure of $P$ = 96-120 GPa) [24].

2. Covalent hydride phosphine, PH₃, with $T_c \simeq 100\,K$ was discovered at $P \gtrsim 200\,GPa$ [25].

3. $PtH_x (x \cong 1)$ recently reported to be superconducting at $P = 30$ GPa [26].

4. $NbTiH_x$ [27].

Unfortunately, we have been unable to consider any of these interesting materials as fundamental experimental data, beyond $T_c$, is unavailable.

This paper shows that all hydrogen-rich superconductors discovered to date, for which experimental data beyond $T_c$ is available, i.e. PdH$_x$, Th$_4$H$_{15}$, Th$_4$D$_{15}$, ThH$_9$, ThH$_{10}$, H$_3$S, and LaH$_{10}$, lie in the same band ($T_c/T_f$ = 0.01-0.05) in the Uemura plot [28-30]. This is the same band as all other unconventional superconductors (heavy fermions, fullerenes, pnictides, and cuprates) – classifying these hydrogen rich compounds as unconventional superconductors. It should be stressed that under some assumptions Th$_4$H$_{15}$ and Th$_4$D$_{15}$ are in closed proximity to Bose-Einstein condensate (BEC) line ($T_c/T_f$ = 0.22) in the Uemura plot.

Here we repeat the analysis described in [6,7], by using the best-known models for upper critical field behaviour we can estimate the ground state coherence length, ξ(0). With this, and the other superconducting parameters, we can calculate the Fermi velocity $v_F$. Then with some knowledge of the effective mass, we can calculate $T_F$ and characterise these conductors in the same manner as Uemura *et al* [28-30].

**II.    The upper critical field models**

The ground state upper critical field, $B_{c2}(0)$, in the Ginzburg-Landau theory [31] is given by:

$$B_{c2}\left(\frac{T}{T_c} = 0\right) = \frac{\phi_0}{2 \cdot \pi \cdot \xi^2(0)}, \qquad (1)$$

where $\phi_0 = 2.068 \cdot 10^{-15}$ Wb is magnetic flux quantum, and ξ(0) is the ground state coherence length. For real world experiments only a part of full $B_{c2}(T)$ temperature dependence can be measured; although there are several models were proposed to deduce extrapolated values for ξ(0) from raw $B_{c2}(T)$ data measured at high reduced temperatures.

One such model, proposed by Werthamer, Helfand, and Hohenberg [32,33], is an extrapolative expression:

$$B_{c2}(0) = \frac{\phi_0}{2\cdot\pi\cdot\xi^2(0)} = -0.693 \cdot T_c \cdot \left(\frac{dB_{c2}(T)}{dT}\right)_{T \sim T_c}, \quad (2)$$

which we designate as the WHH model.

Another model, which is based on the primary idea of the WHH model [32,33], but accurately extrapolates the full $B_{c2}(T)$ curve from experimental data measured at high reduced temperatures, $T/T_c$, was proposed by Baumgartner *et al* [34]:

$$B_{c2}(T) = \frac{\phi_0}{2\cdot\pi\cdot\xi^2(0)} \cdot \left(\frac{\left(1-\frac{T}{T_c}\right) - 0.153\cdot\left(1-\frac{T}{T_c}\right)^2 - 0.152\cdot\left(1-\frac{T}{T_c}\right)^4}{0.693}\right), \quad (3)$$

we will designate this as the B-WHH model.

Gor'kov [35] proposed $B_{c2}(T)$ model which we used in our previous papers [12,13]:

$$B_{c2}(T) = \frac{\phi_0}{2\cdot\pi\cdot\xi^2(0)} \cdot \left(\frac{1.77 - 0.43\cdot\left(\frac{T}{T_c}\right)^2 + 0.07\cdot\left(\frac{T}{T_c}\right)^4}{1.77}\right) \cdot \left[1 - \left(\frac{T}{T_c}\right)^2\right], \quad (4)$$

which we designate as the G-model. Jones *et al* [36], proposed so called Jones-Hulm-Chandrasekhar (JHC) model:

$$B_{c2}(T) = \frac{\phi_0}{2\cdot\pi\cdot\xi^2(0)} \cdot \left(\frac{1-\left(\frac{T}{T_c}\right)^2}{1+\left(\frac{T}{T_c}\right)^2}\right). \quad (5)$$

### III. Th$_4$H$_{15}$-Th$_4$D$_{15}$ superconductors in Uemura plot

We start our consideration with the first discovered superhydride/superdeuteride superconductors i.e. Th$_4$H$_{15}$ and Th$_4$D$_{15}$ [17]. From the author's knowledge, experimental data available to date for the upper critical field, $B_{c2}(T)$, for Th$_4$H$_{15}$ and Th$_4$D$_{15}$ is limited by values reported by Satterthwaite and Toepke [17]. Both $Th_4H_{15}$ and $Th_4D_{15}$ compounds have ground state upper critical fields of:

$$B_{c2}(T\sim0) = 2.5 - 3.0\ T. \tag{6}$$

From these values and Eq. (1), the ground state coherence length, $\xi(0)$, for $Th_4H_{15}$ and $Th_4D_{15}$ phases, must be:

$$\xi(0) = 11.0 \pm 0.5\ nm. \tag{7}$$

Miller *et al* [37] for both phases reported the BCS ratio within a range:

$$\alpha = \frac{2 \cdot \Delta(0)}{k_B \cdot T_c} = 3.42 - 3.47. \tag{8}$$

By using the superconducting transition temperature for $Th_4H_{15}$ and $Th_4D_{15}$ phases [17]:

$$T_c = 8.20 \pm 0.15\ K, \tag{9}$$

one can deduce ground state superconducting energy gap:

$$\Delta(0) = 1.22 \pm 0.03\ meV, \tag{10}$$

and by using well-known BCS expression [10]:

$$\xi(0) = \frac{\hbar \cdot v_F}{\pi \cdot \Delta(0)}, \tag{11}$$

where $\hbar = h/2\pi$ is reduced Planck constant, one can calculate the Fermi velocity, $v_F$, in $Th_4H_{15}$ and $Th_4D_{15}$ phases:

$$v_F = \pi \cdot \frac{\xi(0) \cdot \Delta(0)}{\hbar} = (6.4 \pm 0.2) \cdot 10^4\ m/s. \tag{12}$$

To classify $Th_4H_{15}$ and $Th_4D_{15}$ in the Uemura plot [28-30] we need to make assumption about the effective charge carrier mass, $m^*_{eff}$, to calculate the Fermi temperature, $T_F$:

$$T_F = \frac{\varepsilon_F}{k_B} = \frac{m^*_{eff} \cdot v_F^2}{2 \cdot k_B} \tag{13}$$

As there is no available experimental $m^*_{eff}$ values for $Th_4H_{15}$ and $Th_4D_{15}$ phases, we chose a reasonable lower and upper bound for $m^*_{eff}$. For lower bound we use the value for another ambient pressure hydrogen-rich superconductor, $PdH_x$ [38]:

$$m^*_{eff} = 0.49 \cdot m_e, \tag{14}$$

which leads to the Fermi temperature:

$$T_F = \frac{\varepsilon_F}{k_B} = \frac{m^*_{eff} \cdot v_F^2}{2 \cdot k_B} = 67 \pm 4 \, K, \tag{15}$$

and upper bound to the $T_c/T_f$ ratio:

$$\frac{T_c}{T_F} = 0.12 \pm 0.01. \tag{16}$$

For an upper bound on $m^*_{eff}$ we use the highest value reported for a highly compressed hydrides, $m^*_{eff} = 3.0 \cdot m_e$ [39]. The corresponding lower bound for the $T_c/T_F$ value is then:

$$\frac{T_c}{T_F} = 0.020 \pm 0.002. \tag{17}$$

The above analysis is shown in an Uemura plot, Fig. 1. For one extreme, $m^*_{eff} = 0.49 \cdot m_e$, Th$_4$H$_{15}$ and Th$_4$D$_{15}$ are located in close proximity to Bose-Einstein condensate (BEC) superfluid line, together with $^4$He and $^{40}$K, and thus these two phases cannot be described by BCS theory. For the other bound, $m^*_{eff} = 3.0 \cdot m_e$, Th$_4$H$_{15}$ and Th$_4$D$_{15}$ are still within the band of unconventional superconductors ($T_c/T_f$ = 0.01-0.05) – where all unconventional superconductors (i.e. heavy fermions, fullerens, pnictides and cuprates) are located.

In Figure 1 the BCS ($\frac{T_c}{T_F} \leq 10^{-3}$) and the BEC ($\frac{T_c}{T_F} = 0.22$) boundary lines are plotted to show where all conventional and unconventional superconductors are located.

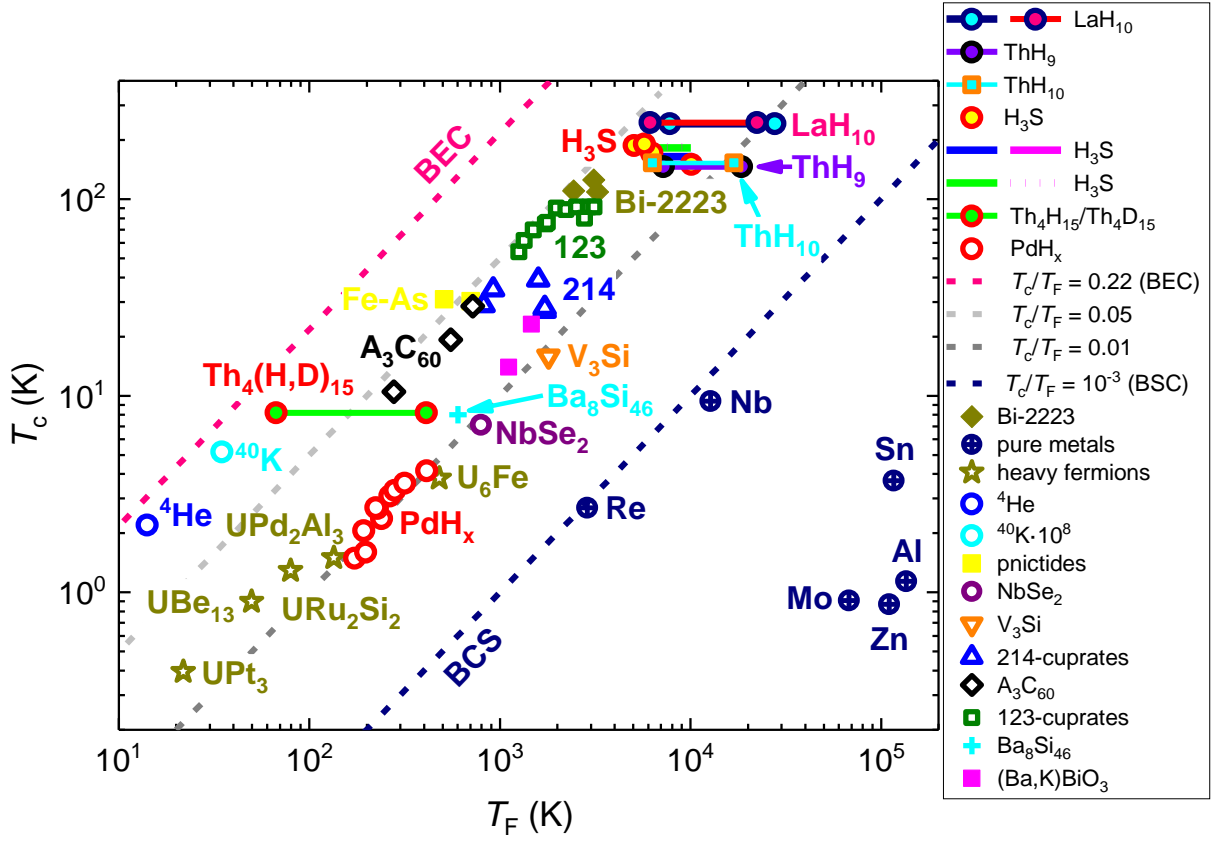

**Figure 1.** A plot of $T_c$ versus $T_F$ obtained for most representative superconducting families including PdH$_x$, Th$_4$H$_{15}$/Th$_4$D$_{15}$, ThH$_9$, ThH$_{10}$, H$_3$S, and LaH$_{10}$. Data was taken from Uemura *et al* [28-30], Ye *et al* [40], Qian *et al* [41], Hashimoto *et al* [42], Shang *et al* [43] and Refs. 12,13. Boundary lines for BEC and BCS materials are shown for clarity.

### IV. ThH$_9$ ($P$ = 170 GPa) in Uemura plot

Semenok *et al* [44] reported the discovery of a high-temperature superconducting phase of ThH$_9$ at $P$ = 170 GPa which exhibits $P6_3/mmc$ crystallographic symmetry and superconducting transition temperature of $T_c$ = 146 K. They also performed first principles calculations and deduced the effective mass in this superconductor:

$$m^*_{eff} = 2.73 \cdot m_e, \qquad (18)$$

which is remarkably close to the effective mass of $m^*_{eff} = 2.76 \cdot m_e$ in compressed H$_3$S [45]. Furthermore, they proposed that ThH$_9$ has BCS ratio:

$$\alpha = \frac{2 \cdot \Delta(0)}{k_B \cdot T_c} = 4.74 - 4.89. \qquad (19)$$

As we mentioned in our previous papers [12,13,46,47], first principles calculations [39,48,49] always provide α-values near 5, which is the very strong-coupling limit for *s*-wave symmetry (also note that other superconducting gap symmetries have weak-coupling limits of α ~ 5 [50-52]).

Despite the orthodox view, several new, alternative, approaches were developed to explain NRT superconductivity in compressed hydrides: Hirsch and Marsiglio [53], Souza and Marsiglio [54], Harshman and Fiory [55], as well as Kaplan and Imry [56]. For instance, Kaplan and Imry [56] showed that in the case of highly compressed $H_3S$ their model gives an α within the weak-coupling BCS limit:

$$\alpha = \frac{2 \cdot \Delta(0)}{k_B \cdot T_c} = 3.53 \qquad (20)$$

This α value is in a good agreement with ones deduced from experimental $B_{c2}(T)$ [12] and the self-field critical current density, $J_c(sf,T)$, data [46,57]. Assuming all hydrogen-rich superconductors have the same primary mechanism for the superconductivity, the value of α = 3.53 was taken as the lower bound for our calculations.

Semenok *et al* [44] measured $B_{c2}(T)$ for $ThH_9$ at $P = 170$ GPa, which we fit to Eqs. 2-5 in Fig. 2. This give $T_c/T_F$ ratios that are within usual range of unconventional superconductors band, See Fig. 1.

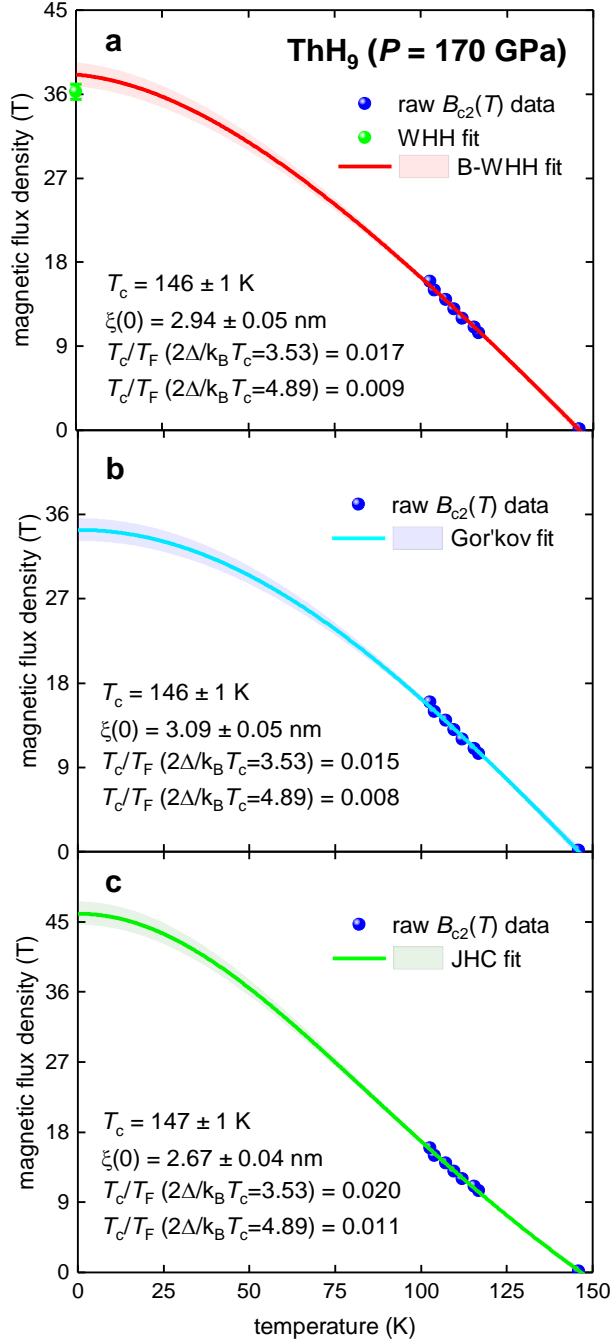

**Figure 2.** Superconducting upper critical field, $B_{c2}(T)$, data and fits to three different models (Eqs. 2-5) for ThH$_9$ superhydride compressed at pressure $P = 170$ GPa (raw data is from [44]). (a) fit to WHH and B-WHH models, for latter the fit quality is $R = 0.998$. (b) fit to Gor'kov model, $R = 0.998$. (c) fit to JHC model, $R = 0.9988$. 95% confidence bars are shown.

### V. ThH$_{10}$ ($P = 174$ GPa) in Uemura plot

Semenok *et al* [44] also reported on the discovery of another high-temperature superconducting phase of ThH$_{10}$ at $P = 174$ GPa, which exhibits $Fm\bar{3}m$ crystallographic symmetry and superconducting transition temperature of $T_c = 159$ K. In Fig. 4 we show raw upper critical field, $B_{c2}(T)$, data for this phase [44] and data fit to Eqs. 2-5.

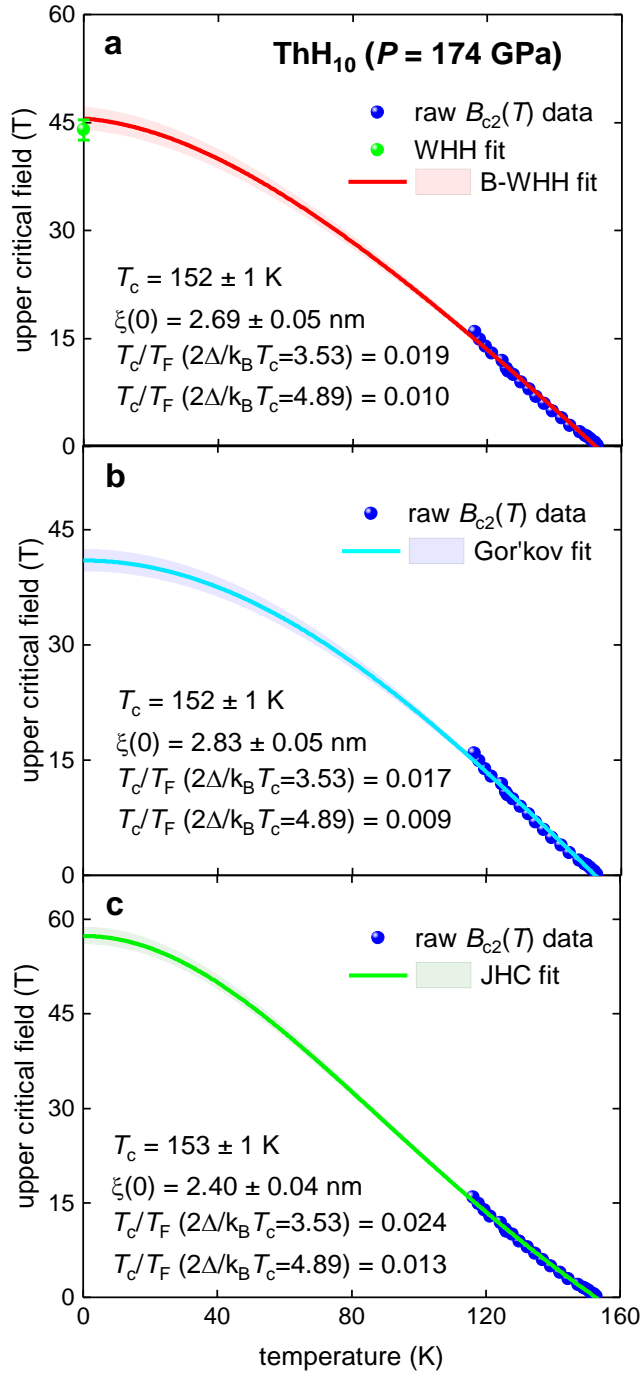

**Figure 3.** Superconducting upper critical field, $B_{c2}(T)$, data and fits to three different models (Eqs. 2-5) for ThH$_{10}$ superhydride compressed at pressure $P = 174$ GPa (raw data is from [44]). (a) fit to WHH and B-WHH models, for latter the fit quality is $R = 0.992$. (b) fit to Gor'kov model, $R = 0.992$. (c) fit to JHC model, $R = 0.997$. 95% confidence bars are shown.

As expected, highly-compressed ThH$_{10}$ superconductor is located within unconventional superconductors band of the Uemura plot, see Fig. 2.

## VI. Conclusions

Recent interest in the near-room-temperature superconductivity has revived interest in the hydride superconductors. While the latest generation of hydride superconductors, H$_3$S-D$_3$S and LaH$_{10}$-LaD$_{11}$/LaD$_{12}$, are widely considered to be conventional BCS conductors, we point out that this is not supported in other hydrides such as the Th$_4$H$_{15}$-Th$_4$D$_{15}$, and PdH$_x$-PdD$_x$-PdT$_x$. Critically, these previously discovered hydride systems exhibit the reverse isotope effect, which cannot be explained in BCS theory. In addition, we stress that the isotope effect in LaH-LaD system should be further studied, as available experimental data show that at high-pressure conditions La-H and La-D NRT superconducting phases have different stoichiometry [23].

To further this analysis, we have classified (conventional vs unconventional) the superconductivity in the thorium hydrides. We analyse experimental $B_{c2}(T)$ data for several thorium based superhydrides and Th$_4$D$_{15}$ superdeuteride. This analysis was completed for thorium hydrides where fundamental superconducting parameters beyond $T_c$ were available i.e., Th$_4$H$_{15}$, Th$_4$D$_{15}$, ThH$_9$ and ThH$_{10}$. For all these materials – all thorium hydrides where analysis is possible – we find that they fall into the band of unconventional superconductors, as seen in an Uemura plot. This along with similar analysis of other hydrides, previously done, further necessitates understanding the hydrides outside of conventional BCS theory.

### Acknowledgement

Author thanks financial support provided by the state assignment of Minobrnauki of Russia (theme "Pressure" No. AAAA-A18-118020190104-3) and by Act 211 Government of the Russian Federation, contract No. 02.A03.21.0006.